\documentclass[11pt]{jaa}
\usepackage[utf8]{inputenc}
\usepackage[T1]{fontenc}
\usepackage{graphicx}
\usepackage{longtable}
\usepackage{wrapfig}
\usepackage{rotating}
\usepackage[normalem]{ulem}
\usepackage{amsmath}
\usepackage{amssymb}
\usepackage{capt-of}
\usepackage{hyperref}
\usepackage{natbib}
\usepackage{lipsum}
\renewcommand{\thesection}{\arabic{section}}
\renewcommand{\thesubsection}{\thesection.\arabic{subsection}}

\hypersetup{colorlinks=true,linkcolor=blue,citecolor=blue,urlcolor=blue}
\affilOne{\textsuperscript{1}Tata Institute of Fundamental Research, Homi Bhabha Road, Colaba, Mumbai 400 005, India.\\}
\affilTwo{\textsuperscript{2}University of Hertfordshire Hatfield, AL10 9AB, United Kingdom\\}
\affilThree{\textsuperscript{3}Aryabhatta Research Institute of Observational Sciences, Manora Peak, Nainital-263001, Uttarakhand, India.\\}

\author{Varghese Reji\textsuperscript{1}\textsuperscript{*}, Joe P. Ninan\textsuperscript{1}, Supriyo Ghosh\textsuperscript{2}, Devendra K. Ojha\textsuperscript{1}, Saurabh Sharma\textsuperscript{3}}
\date{}
\title{\texttt{pyTANSPEC-v1.0} and \texttt{HxRGproc}: Updated packages to Clean and Reduce TANSPEC data}
\hypersetup{
 pdfauthor={Varghese Reji\textsuperscript{1}\textsuperscript{*}, Joe Ninan\textsuperscript{1}, Supriyo Ghosh\textsuperscript{2}, D.K. Ojha\textsuperscript{1}, Saurabh Sharma\textsuperscript{3}},
 pdftitle={pyTANSPEC v1.0 and HxRGproc: Updated packages to Clean and Reduce TANSPEC data},
 pdfkeywords={},
 pdfsubject={},
 pdfcreator={Emacs 28.2 (Org mode 9.6.7)}, 
 pdflang={English}}
\begin{document}

\twocolumn[{
\maketitle
\corres{varghesereji0007@gmail.com}

\begin{abstract}
TIFR-ARIES Near-Infrared Spectrometer (TANSPEC) is a spectrograph-cum-imager operating over the wavelength range $0.55 - 2.5~\mu$m. The instrument is mounted on the 3.6-m Devasthal Optical Telescope  (3.6-m DOT). It offers two resolution modes: Low Resolution (LR) with $R\sim100-350$ and Cross-Dispersed (XD) {via various slits of different widths (0.5", 0.75", 1.0", 1.5", 2.0" and 4.0"). The LR mode provides a resolving power ($R$) of $\sim 100-350$, while the XD mode achieves $R\sim2500$ using the 0.5" slit.} The previous version {of the data reduction pipeline supported only wavelength-calibrated XD mode spectra and was limited to two slits (S-0.5 and S-1.0).} In this work, we present an upgraded version of \texttt{pyTANSPEC}. {The upgraded pipeline not only improves the data extraction algorithm but also introduces several new features for users. It now enables the reduction of spectra from all available slits for both LR and XD modes.  
The upgraded version also implements a template-matching method for more precise wavelength calibration. Additionally, a step for flux calibration is also included.} 

Alongside \texttt{pyTANSPEC}, we upgraded \texttt{HxRGproc}, a Python package for cleaning and generating slope images from Non-Destructive Readout (NDR) frames taken with H1RG and H2RG detectors. The package performs non-linearity correction, flags saturated pixels, removes pink noise, and eliminates cosmic ray events. \texttt{HxRGproc} {is updated to work for the H2RG detector of TANSPEC and} is set up on the TANSPEC server, ensuring users receive data that are pre-cleaned and non-linearity corrected.
\end{abstract}
}]

\doinum{00.0000/0000-000-000-0}
\artcitid{\#\#\#\#}
\volnum{000}
\year{0000}

\section{Introduction}
\label{sec:org4bb5a90}
TANSPEC (TIFR-ARIES Near-Infrared Spectrometer) is a spectrograph-cum-imager \citep{2022} that has been operational on the 3.6-m Devasthal Optical Telescope (3.6-m DOT) since 2019. TANSPEC is a unique instrument across the world with its broad wavelength coverage ($0.55-2.5~\mu$m). {It has two spectral resolution modes with resolutions of $\sim 2500$ in High Resolution or Cross Dispersion (XD) mode, and $\sim 100-350$ in Low Resolution (LR) mode. In addition, TANSPEC can also do imaging}. In its spectrograph arm, TANSPEC uses an H2RG detector \citep{2011ASPC..437..383B} with four readout channels. The H2RG detector is an array of infrared light-sensitive diodes, which are hybridised to CMOS MOSFET-switched multiplexers \citep{FINGER2006241}.

{TANSPEC is always operated in sample-up-the-ramp (SUTR) mode}. Data obtained with H2RG detectors exhibit significant non-Gaussian spatial and temporal noise, such as classical non-linearity \citep{SMADJA2009615}, bias fluctuations, persistence and other effects. Some of these noise sources are corrected during pre-processing. \texttt{HxRGproc} is a Python package to pre-process the raw Non-Destructive Readout (NDR) frames from H2RG and generate slope images \footnote{During an exposure, the absolute counts in each pixel are read out and saved at equal time intervals. These frames are called Non-Destructive Readout (NDR) frames because the readout process does not reset the pixel to its initial value. After the exposure, the software fits a straight line to the NDR values of each pixel, and the slope of this fit is stored. The resulting image is known as the slope image.} \citep{2018SPIE10709E..2UN, ninan_indiajoehxrgproc_2021}. {The pre-processing includes cosmic-ray correction, correction on bias fluctuation, and non-linearity correction.} The package was originally made for the Habitable-zone Planet Finder (HPF) spectrograph \citep{Mahadevan_2012, Mahadevan_2014}. As a part of this work, the package was upgraded to support the data taken by the spectrograph arm of TANSPEC{, which is using the H2RG detector}. In this manuscript, the 2D slope image data after pre-processing with \texttt{HxRGproc} is referred to as an HxRGproc frame, and the 2D frame generated by TANSPEC software at the time of observation is referred to as a Quick Look Frame (QLF).

\texttt{pyTANSPEC} reduces and extract the spectra from 2D slope images. \texttt{pyTANSPEC v0.0.1} \citep{supriyo22} supported reduction of XD mode spectra taken with TANSPEC. However, the low-resolution (LR) mode of TANSPEC is also widely used by the community for studies of faint objects, spectral energy distributions, transmission spectroscopy of exoplanets, etc. Therefore, this upgrade enables observers to fully utilise the LR spectrograph mode of TANSPEC.

\autoref{sec:org1a6af4b} discusses bias correction and non-linearity correction of data using HxRGproc. \autoref{sec:orgbce9730} discusses detector performance like readout noise and gain. \autoref{sec:orgee7a7f3} discusses the upgrade of \texttt{pyTANSPEC}, the methodology of wavelength and flux calibration, etc. \autoref{sec:org5ae93e5} discusses the performance of the pipeline in the test data and outlines plans for the pipeline.

\section{Pre-processing with HxRGproc}
\label{sec:org1a6af4b}

\subsection{Correction of Bias Fluctuations}
\label{sec:org4ec3eb4}
Bias fluctuation during readout is one of the sources of non-Gaussian errors in HxRG detectors. It is additive in nature, adding to the pixel readout before digitisation by the read-out integrated circuit (ROIC), and it exhibits 1/f (flicker) noise, i.e. noise whose power spectral density scales inversely with frequency \citep{Rauscher_2015}. Reference pixels, located in the first and last four columns and rows of HxRG detectors, do not capture light \citep{Rauscher_2017}. Therefore, the counts in these pixels contain bias fluctuation and readout noise.

\begin{figure*}[htbp]
\centering
\includegraphics[width=.9\linewidth]{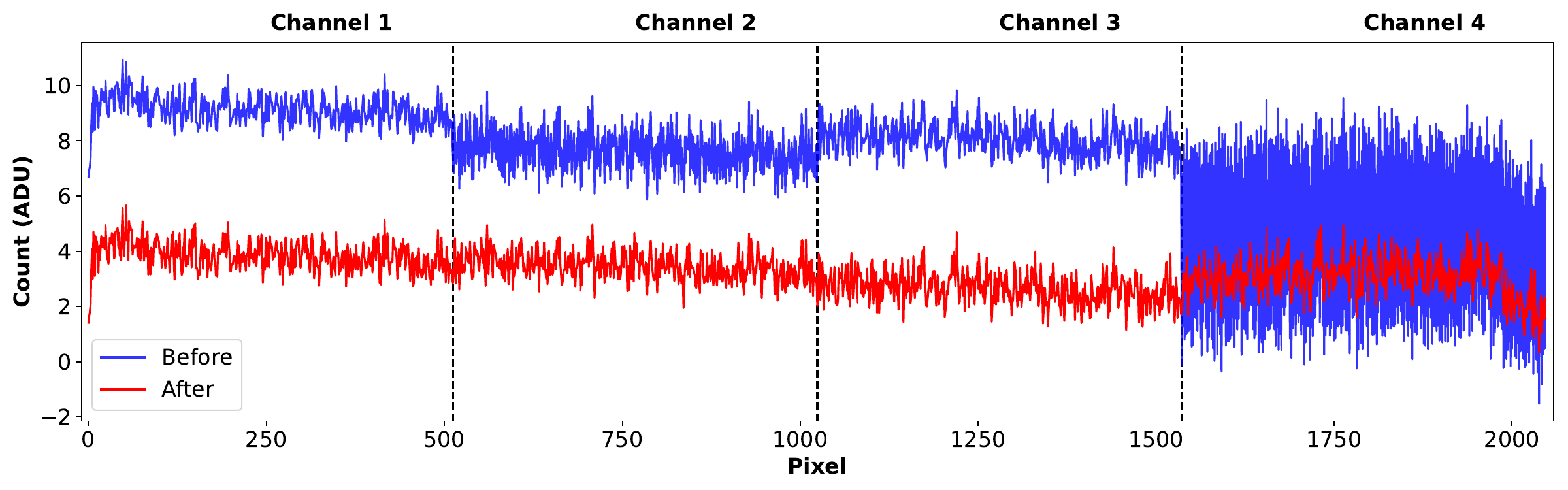}\\
\includegraphics[width=.9\linewidth]{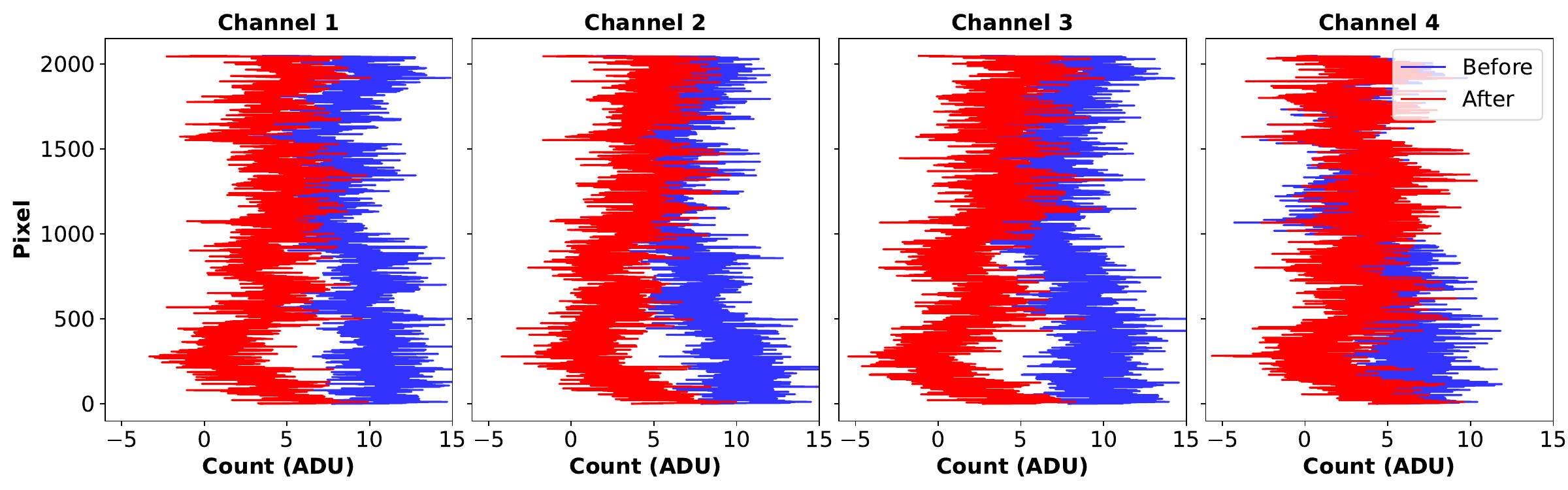}
\caption{\label{fig:bias_noise}Comparison of the bias fluctuation in the NDR frames {of H2RG detector of TANSPEC} before and after bias correction in both horizontal (top panel) and vertical (bottom panel) directions. Blue represents the bias fluctuation in the raw data, while red shows the fluctuation after bias correction with \texttt{HxRGproc}. In the top plot, counts are averaged along the vertical direction to average out readout noise. Similarly, in the bottom plot, counts are averaged along the horizontal direction. The plots illustrate the bias fluctuations in both directions. There is a notable improvement in bias fluctuation in the horizontal direction, with the count level decreasing from approximately 9 to 4 after bias correction. In contrast, no significant difference is observed in the bias noise in the vertical direction across individual readout channels.}
\end{figure*}

\texttt{HxRGproc} uses the counts of reference pixels to subtract the bias in each NDR frame. However, the reference pixels themselves are subject to detector readout noise; consequently, their direct subtraction increases the effective readout noise. To address this, the counts in the reference pixels are smoothed with a Savitzky-Golay (SG) filter using a second-degree polynomial \citep{sg_filter}. The optimal bias correction for TANSPEC was achieved with an SG filter window size of 439 pixels. {This window size was optimised to minimise the effective readout noise after bias correction in dark frames}. A comparison of bias fluctuation before and after correction in both horizontal and vertical directions is shown in \autoref{fig:bias_noise}. The figure demonstrates a clear improvement in horizontal noise after correction, most notably in the suppression of alternating column noise in Channel 4. The horizontal noise is effectively reduced by a factor of  $\sim3.4$. In contrast, the vertical noise remains largely unchanged, with a reduction factor of only 1.03.

\subsection{Correction for Non-linearity}
\label{sec:orga038dfb}
The non-linear relationship between the incident photon count and the electrical output signal is referred to as non-linearity in a detector. This is a major issue that affects photometry and spectroscopy of bright sources. Non-linearity correction can be performed on H2RG NDR frames following the algorithm described in  \cite{2018SPIE10709E..2UN}, implemented via \texttt{HxRGproc}.

\begin{figure}[htbp]
\centering
\includegraphics[width=\linewidth]{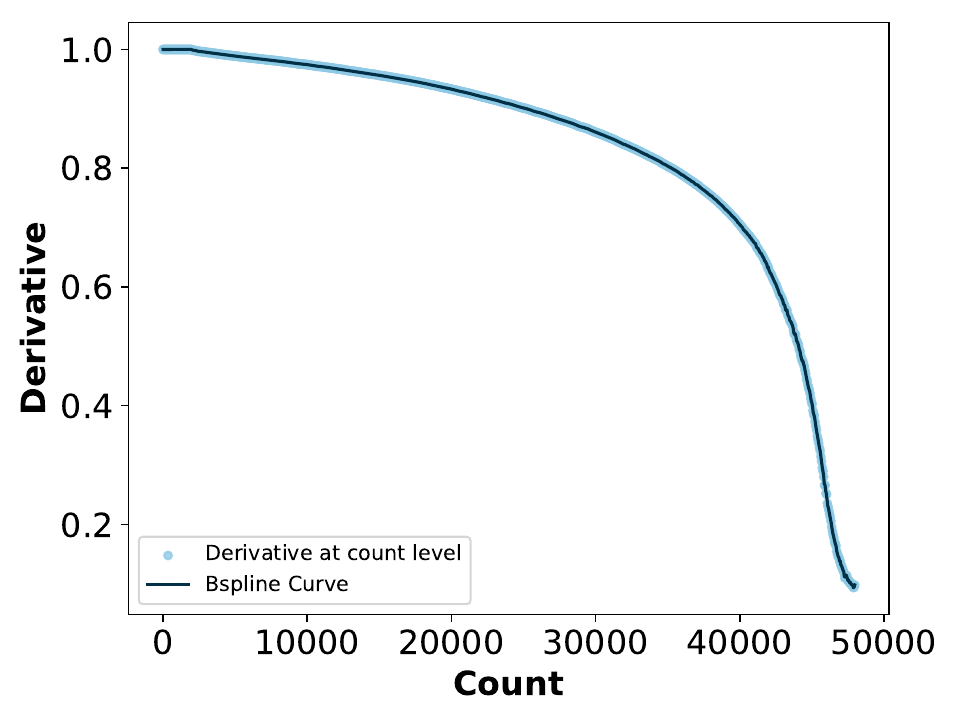}
\caption{\label{fig:orgc64a36c}Non-linearity curve of the TANSPEC H2RG detector. Light blue dots represent the binned and normalised derivative with respect to counts. The black curve shows the B-spline fit used for integration over count levels. This integral provides the corrected value.}
\end{figure}

For a detector, the change in count is proportional to the incoming flux. \begin{equation}
    \frac{dC}{dt} \propto F \implies \frac{dC}{dt} = \alpha F 
\end{equation}
where $C$ is the electron count and $F$ is the incoming flux. $\alpha$ is the constant of proportionality. However, in real detectors, $\alpha$ will be a function of count $C$. The solution of the differential equation is:
\begin{equation}
    FT = \int_0^C \frac{dC'}{\alpha(C')}
\end{equation}
where $T$ is the exposure time. The right-hand side of the above equation is a function of count $C$. To obtain this calibration curve, the detector was exposed to the continuum light through the slit of width 1.5" and length 60" (L-1.5) in XD mode, ensuring illumination of the majority of the pixels in the detector. {The readout was taken in High-Gain (HG) setting of the detector \citep{2022}. \footnote{TANSPEC is currently operated only in the HG setting.}}To reduce bias and readout noise, bias-corrected and averaged NDR frames were then created. The derivative of counts was calculated by differencing successive averaged NDR frames. Pixels were binned by the count level, and the corresponding count derivatives were selected. Data modelling with the inverse B-spline non-linearity model is explained in \cite{2018SPIE10709E..2UN}. The inverse response function was fitted using a cubic B-spline, with knot placement and number determined automatically by the FITPACK algorithm \citep{10.1093/oso/9780198534419.001.0001} implemented in \texttt{scipy.interpolate.splrep}. The fitted B-spline curve is shown as the blue curve in \autoref{fig:orgc64a36c}.

\begin{figure*}[htbp]
\centering
\includegraphics[width=\linewidth]{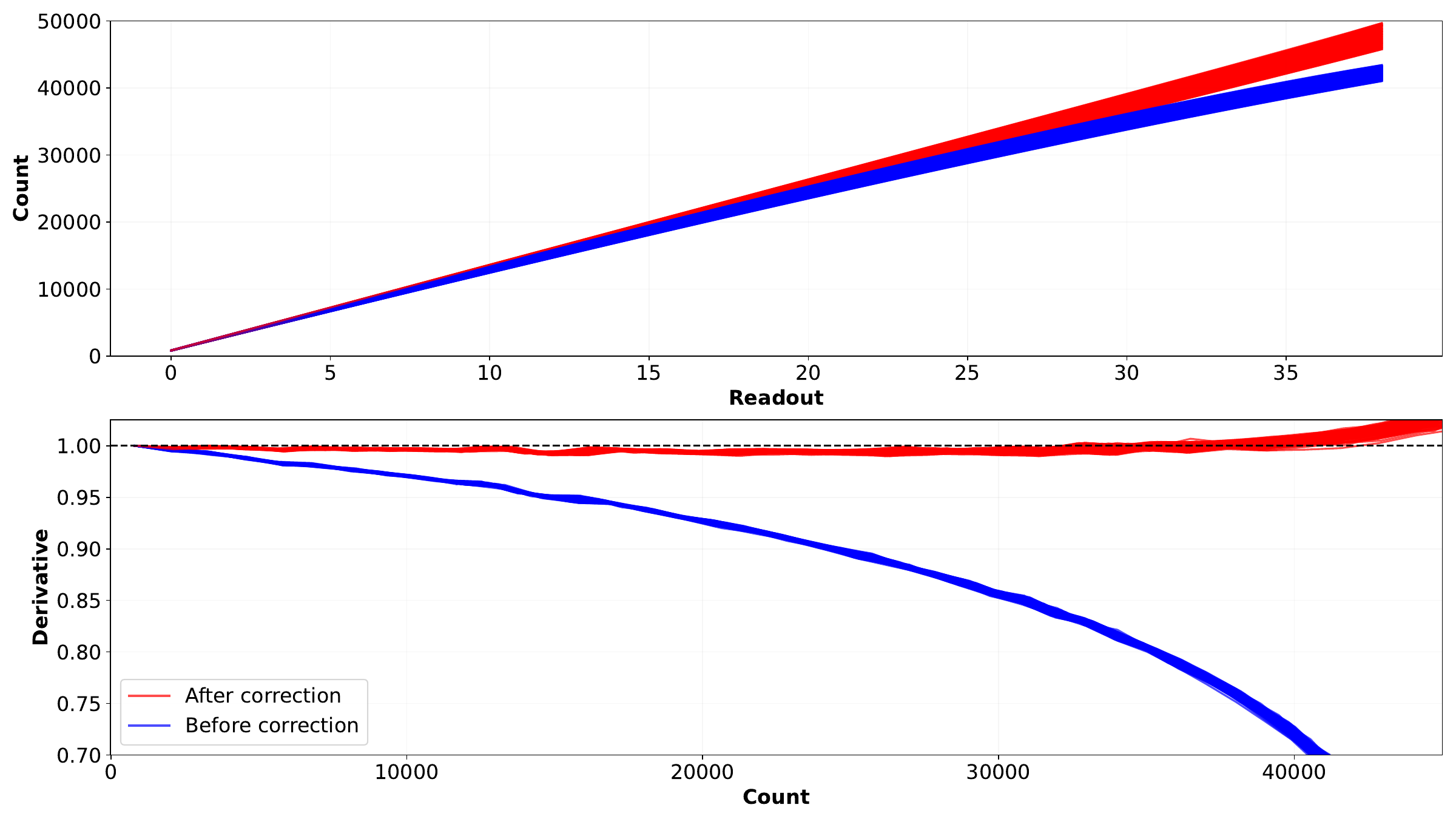}
\caption{\label{fig:comp_nlty} Top panel: Comparison of up-the-ramp counts from NDR frames before and after application of non-linearity correction. Bottom panel: The difference between successive frames normalised by the difference between the first two frames. The difference between successive frames becomes constant after the non-linearity correction.}
\end{figure*}

\begin{figure*}[t!]
\centering
\includegraphics[width=\textwidth]{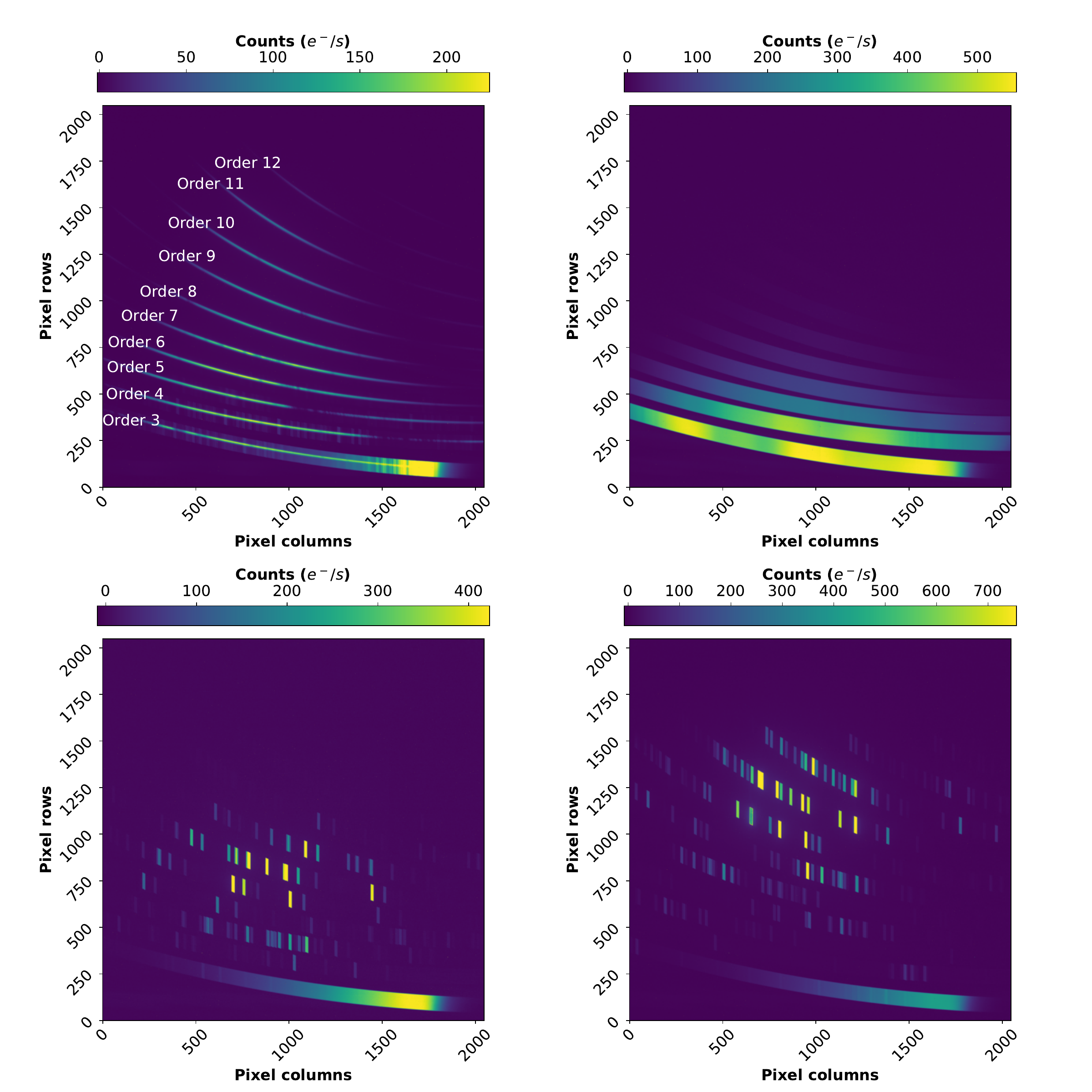}
\caption{\label{fig:xd_raw}Raw spectral images in XD mode: science frame (top-left), continuum flat (top-right), Argon lamp (bottom-left), and Neon lamp (bottom-right). The colour bar indicates counts in e\textsuperscript{-}/s. Orders 3($\lambda \sim 1.9{-}2.5$ $\mu$m) to 12 ($\lambda \sim 0.55{-}0.61$ $\mu$m) are visible from bottom to top, with shorter wavelengths on the left and longer wavelengths on the right of each order. Thermal continuum emission from warm optics outside the cryostat is seen in the K band.}
\end{figure*}

\begin{figure*}[t!]
\centering
\includegraphics[width=\textwidth]{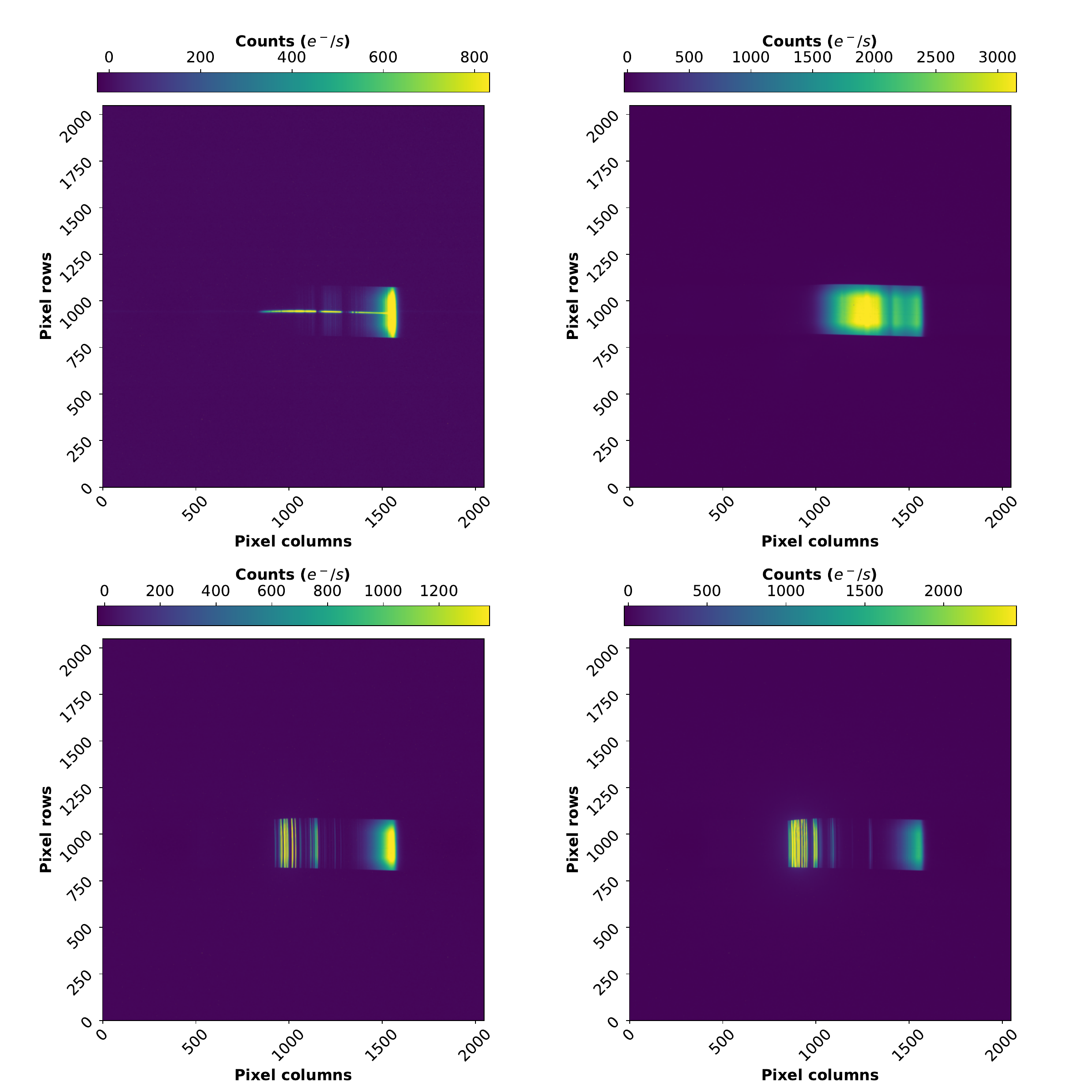}
\caption{\label{fig:lr_raw} Raw spectral images in LR mode: science frame (top-left), continuum flat (top-right), Argon lamp (bottom-left), and Neon lamp (bottom-right). The colour bar indicates counts. The spectra span a wavelength range of $\sim 0.55{-}2.5$ $\mu$m, with the shortest wavelengths on the left-hand side of the aperture and the longest on the right-hand side.}
\end{figure*}

\begin{figure*}[htbp]
\centering
\includegraphics[width=.9\linewidth]{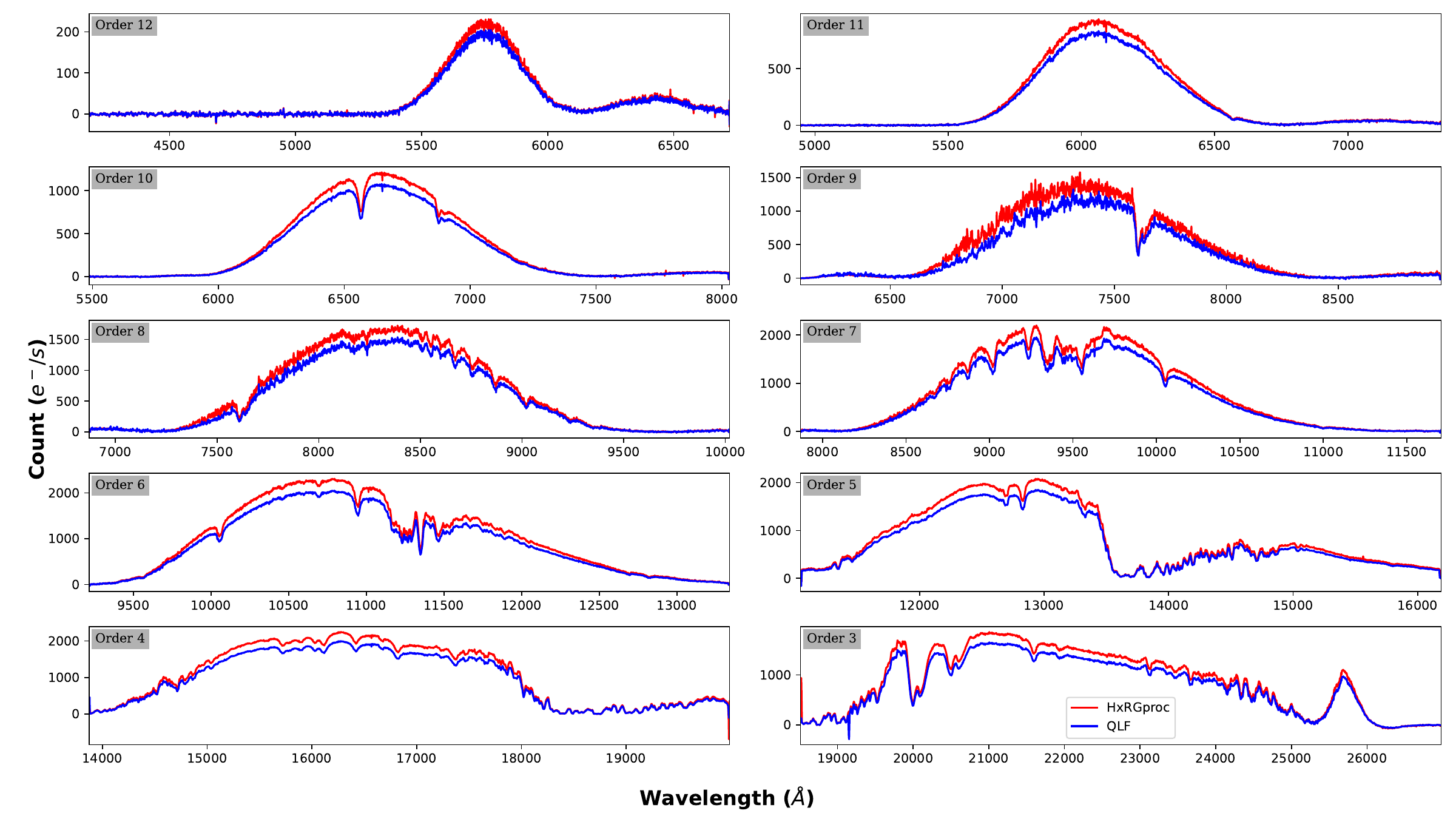}
\caption{\label{fig:hxrg_qlf}Comparison of spectra of the same star extracted from the QLF (blue) and from \texttt{HxRGproc} (red) frames. The spectra from \texttt{HxRGproc} show higher flux levels due to the improvement from the non-linearity correction applied during processing. After applying the non-linearity correction, the extracted flux increases by approximately 13\%.}
\end{figure*}

A comparison of the NDR data before and after non-linearity correction is shown in  \autoref{fig:comp_nlty}. After correction, the derivative remains nearly constant and shows a clear improvement over the uncorrected case. However, it begins to deviate from the constant behaviour beyond a count level of \(\sim35,000\) ADU ($\sim 37400$ e\textsuperscript{-}). Consequently, we set 35,000 ADU as the effective saturation threshold above which data cannot be recovered, and \texttt{HxRGproc} excludes pixels that exceed this threshold when generating the slope image. The 2D spectral images obtained with \texttt{HxRGproc} after bias and non-linearity correction are shown in \autoref{fig:xd_raw} (XD mode) and in \autoref{fig:lr_raw} (LR mode). \autoref{fig:hxrg_qlf} compares spectra extracted from the QLF and \texttt{HxRGproc} frames, showing an increase of approximately 13\% in flux after non-linearity correction.

\subsection{Correction of Cosmic Ray Events}
{\texttt{HxRGproc} implements the cosmic ray (CR) correction algorithm described in \cite{ninan14:_tirsp} for the correction of CR events. In the SUTR mode, the time-resolved signal exhibits a sudden discontinuity at the pixel when a CR event occurs. Sudden discontinuities are identified using a digital filter constructed by convolving the first difference filter and a Mexican hat filter. Once a CR event is detected, the signal slope is estimated for the segments preceding and following the discontinuity. The final slope of the counts of that pixel is calculated by a weighted average of these slopes.}

\section{Detector Performance After Pre-processing}
\label{sec:orgbce9730}
\subsection{Readout noise}
\label{sec:orge6d3667}
Readout noise is the random noise added by the amplifier circuit to the signal during readout. \cite{2022} reported the readout noise of the TANSPEC H2RG detector. Here, we re-calculate the readout noise from the NDR frames after improved bias fluctuation correction. To estimate readout noise, we acquired 21 dark exposures, each consisting of 60 readouts. We computed the difference between consecutive frames along the time axis (axis=0 of the datacube) to isolate the noise component. From this, we selected every second frame, and then calculated the variance along the time axis, dividing it by two to account for the differencing process, and finally took the square root of this result to obtain the readout noise values. This procedure yields the distribution of {effective readout noise.} We calculated the {effective readout noise} before and after bias correction and observed only minor improvement.

\begin{figure*}[htbp]
\centering
\includegraphics[width=.9\linewidth]{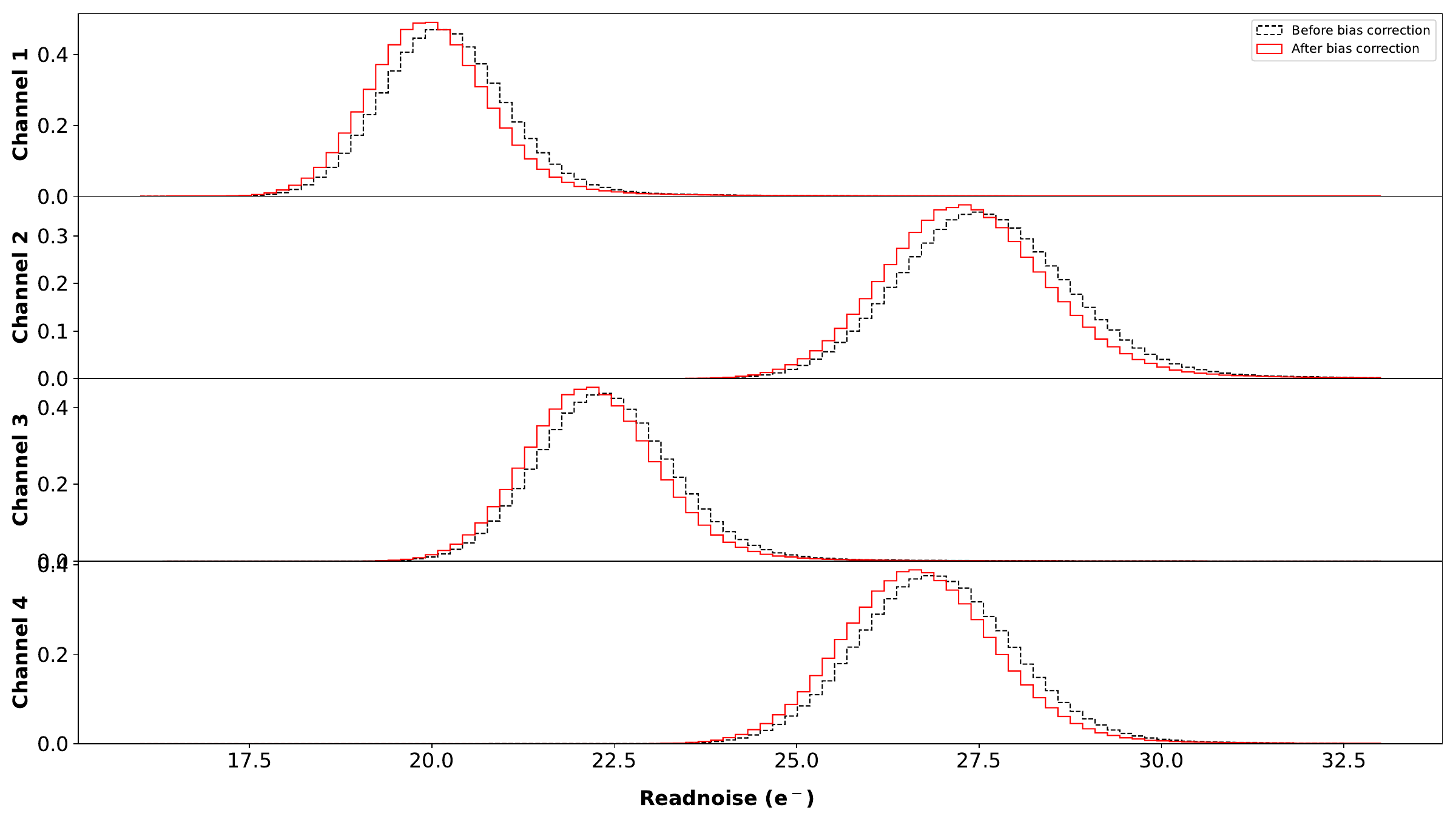}
\caption{\label{fig:read_noise} Histogram of readout noise across channels before (black) and after (red) bias correction. }
\end{figure*}

Our analysis shows that the readout noise varies across different readout channels. This is because each channel is connected to a different amplifier, which results in channel-wise variations in detector behaviour. The channel-wise summary of the readout noise with and without bias correction is shown in \autoref{tab:orge946e56}, and the distribution of the readout noise is shown in \autoref{fig:read_noise}.

\begin{table*}[htbp]

\centering
\begin{tabular}{cccc}
\hline
Channel & \multicolumn{2}{c}{{Effective} readout noise (e\textsuperscript{-})}  & Gain (e\textsuperscript{-}/ADU) \\[0pt]
 & Before bias correction & After bias correction & \\[0pt]
\hline

1 & $20.2 \pm 3.8$ & $20.0 \pm 3.8$ &  $1.07\pm0.10$ \\[0pt]
2 & $27.6 \pm 6.0$ & $27.3 \pm 5.9$ & $1.37\pm0.15$ \\[0pt]
3 & $22.4 \pm 4.0$ & $22.2 \pm 4.0$ & $1.00\pm0.08$ \\[0pt]
4 & $26.9 \pm 4.7$ & $26.7 \pm 4.7$ & $1.07\pm0.11$ \\[0pt]
\hline
Median & $24.6\pm3.1$ & $24.4\pm3.1$ & $1.07\pm0.07$ \\[0pt]
\hline
\end{tabular}
\caption{\label{tab:orge946e56}Summary of the readout noise in each channel, with and without bias correction. The channel-wise gain values are also included in the table.}
\end{table*}

\begin{figure}
    \centering
    \includegraphics[width=\linewidth]{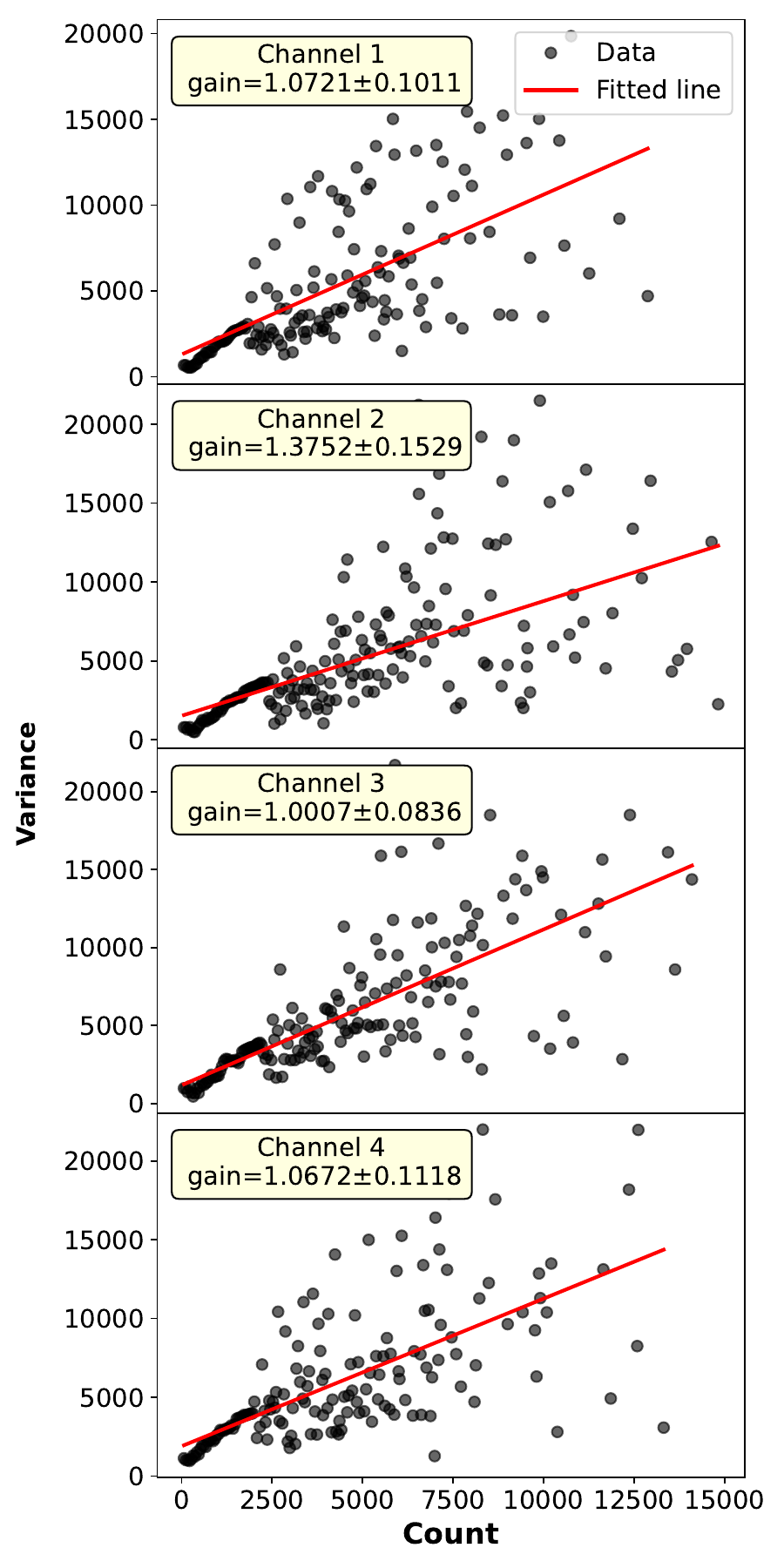}
    \caption{Channel-wise photon transfer curve of the H2RG detector of TANSPEC. Black dots are the data points, and the red line is the fitted curve. The inverse of the slope of the line is the gain.}
    \label{fig:gain}
\end{figure}

\subsection{Gain}
\label{sec:orgc527e44}
\cite{2022} reported the gain in the high gain mode of the H2RG detector to be approximately \(1.12\pm 0.03\) e\textsuperscript{-}/ADU. Here, we re-calculate the gain in the high-gain mode for each channel separately using the mean-variance technique, also known as the photon transfer technique. For this, we acquired 21 sets of exposures, each consisting of four readouts of continuum light through slit L-1.5 in XD mode, ensuring maximum coverage of the pixels in the detector. After applying non-linearity and bias corrections, the mean and variance of each readout across exposures were calculated. The variance values were binned by count level, and a straight line was fitted to the data. The inverse of the slope of this line gives the gain for each channel. The photon transfer curves per channel are shown in \autoref{fig:gain}. The calculated gain values are summarised in Table \ref{tab:orge946e56}.

\section{Spectral extraction with pyTANSPEC}
\label{sec:orgee7a7f3}

{The earlier pipeline lacked LR extraction support because of differences in the tracing algorithms.} The major upgrade to \texttt{pyTANSPEC-v1.0} is to extract LR- and XD-mode spectra and perform wavelength calibration for TANSPEC data in a self-consistent manner. In addition to the tasks available in the earlier version \citep{supriyo22}, an additional task (Task 7) for flux calibration has also been incorporated. The new procedures for spectral extraction, wavelength calibration, and flux calibration are described in the following subsections.

\subsection{Extraction of Spectra}
\label{sec:org902141b}
Spectral extraction in LR mode follows the same algorithm as in XD mode, differing only in aperture. The package \texttt{SpectrumExtractor} \footnote{\url{https://github.com/indiajoe/SpectrumExtractor.git}} extracts target spectra from 2D spectral images. The curvature in apertures makes tracing the aperture position in the image difficult. Therefore, a trace was created using \texttt{SpectrumExtractor} for both spectral modes and saved with the pipeline. XD mode has 10 apertures for its spectral orders, whereas LR mode has only one, so their trace files differ.

LR trace was newly created from the slope images generated with \texttt{HxRGproc} and are included with the upgraded pipeline. We also updated the XD mode trace in this upgrade. These traces are used to locate the initial guess of apertures during extraction, based on the LR/XD choice set in the configuration file. Then, the traces were fitted with the frame using a polynomial. \cite{supriyo22} were using a first-order polynomial for trace fitting. In the current version, trace fitting is performed using a second-order polynomial, which provides increased flexibility in capturing curvature across the aperture.

A new feature overlays the fitted aperture traces on the 2D image for visual inspection and allows adjustments to the aperture fitting parameters, such as aperture window and background window, using the config file. Users may also create and use custom trace files.

\subsection{Wavelength Calibration}
\label{sec:org398eec2}
The \texttt{WavelengthCalibrationTool} \footnote{\url{https://github.com/indiajoe/WavelengthCalibrationTool.git}} is used for wavelength calibration. In \texttt{pyTANSPEC v0.0.1} \citep{supriyo22}, this was done by \textit{re-identifying} the pixel positions of non-blended argon and neon emission lines from lamp spectra taken after the science frames. During data reduction, the pipeline automatically re-identified these non-blended lines in the observed lamp spectra and fitted a polynomial to the pixel–wavelength relation.

\begin{figure*}
    \centering
    \includegraphics[width=\linewidth]{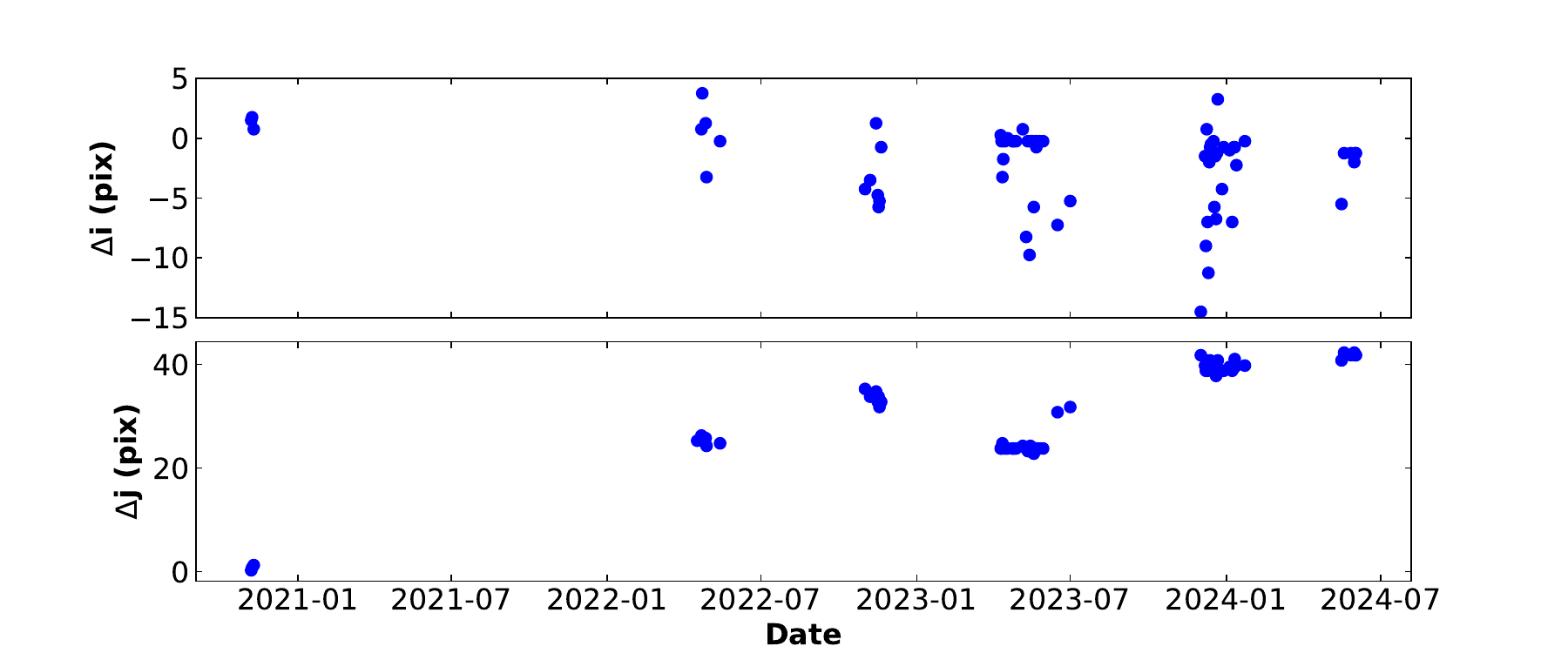}
    \caption{Detector pixel position shifts measured over time, derived from argon and neon lamp frames acquired between 2020 November and 2024 June. The shifts were estimated using phase cross-correlation. The top panel shows the shift in the vertical direction, while the bottom panel shows the shift in the horizontal direction. The magnitude of these shifts is sufficient to affect the performance of the automatic re-identification algorithm implemented in the previous version of \texttt{pyTANSPEC}.}
    \label{fig:pixshift}
\end{figure*}

However, this method has two key limitations:
\begin{enumerate}
\item For wider slits (other than 0.5") and LR spectra, spectral lines are often blended, making centroid determination difficult.
\item Aperture position on the detector can shift significantly between observation cycles at DOT, which can cause the automatic re-identification algorithm to fail. This became clear after a few years of operations. The shift in pixel position calculated using argon and neon lamps taken between 2020 November and 2024 June is shown in \autoref{fig:pixshift}.
\end{enumerate}

To address these issues, the new version uses \textit{template matching}, {by least squares minimisation,} where observed lamp spectra are matched to pre-calibrated template spectra. This approach applies to all slits in both LR and XD modes.

\begin{figure*}[htbp]
\centering
\includegraphics[width=\linewidth]{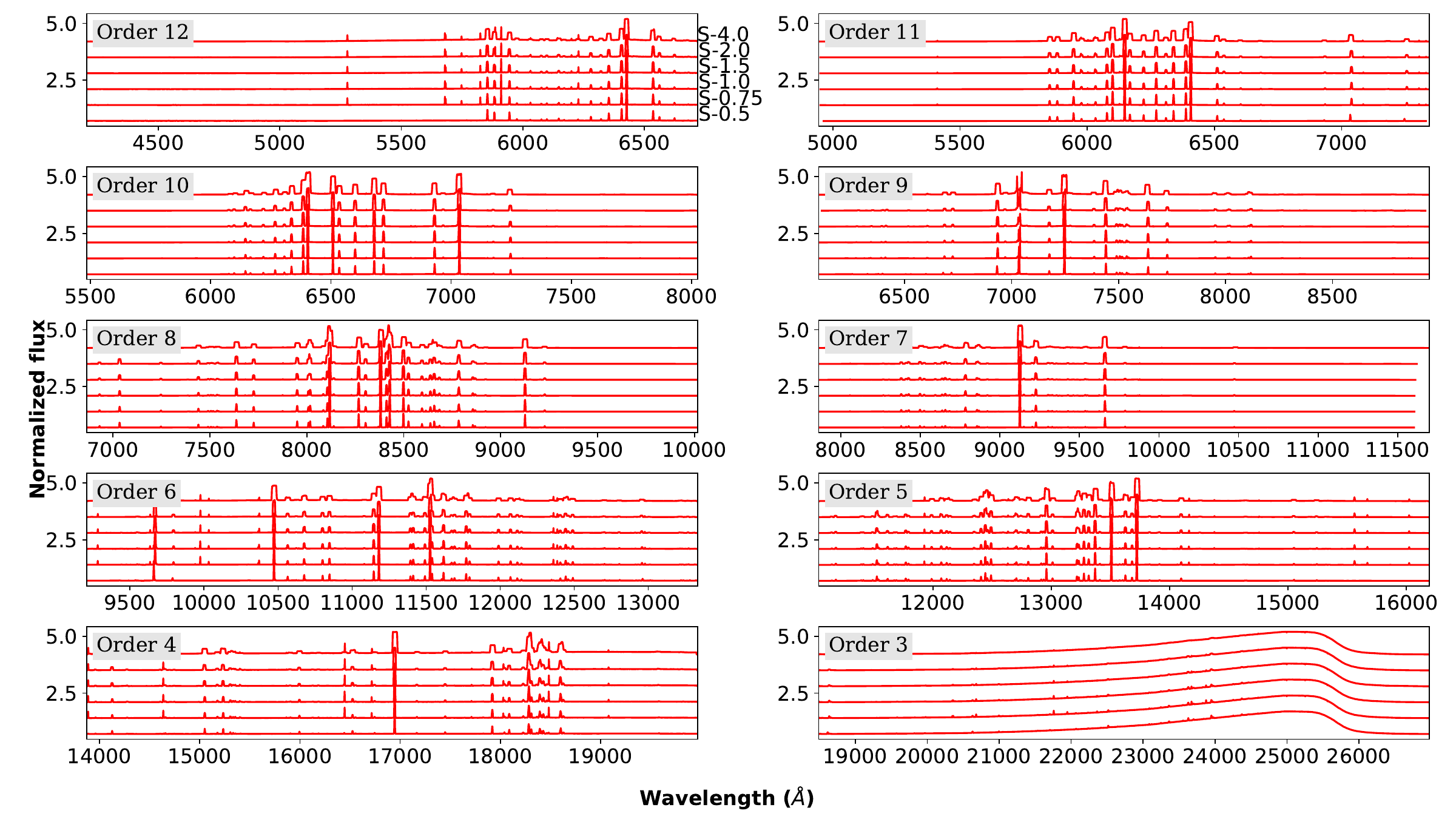}
\includegraphics[width=\linewidth]{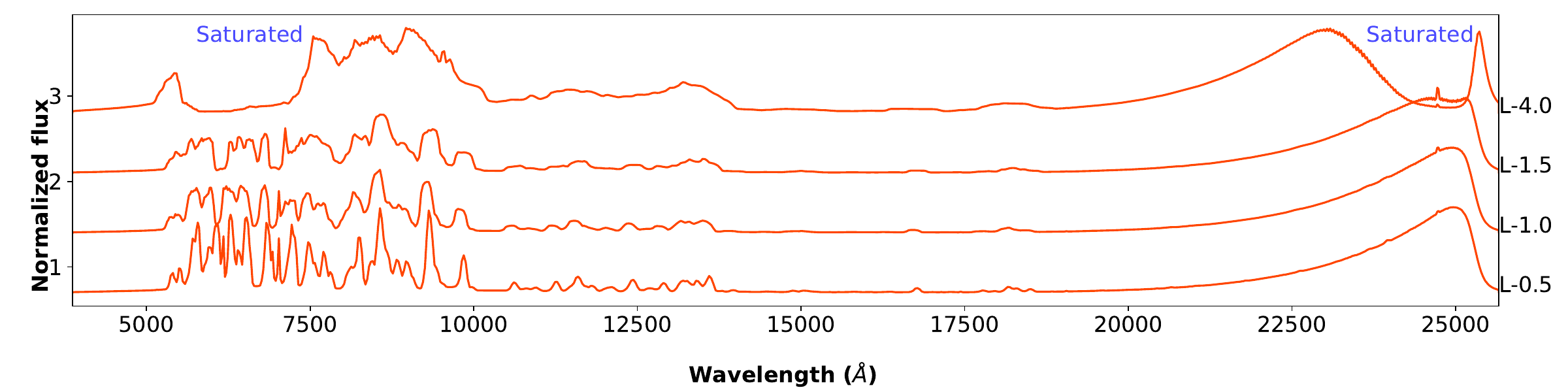}
\caption{\label{fig:xd_temp} Template spectra for wavelength calibration. Top panel: XD mode templates for all spectral orders. Bottom panel: LR mode templates.}
\end{figure*}

\textbf{Template creation process:} We obtained 21 exposures of argon and neon lamps, each with an exposure time of 25 s, for all slits. Slope images were generated using \texttt{HxRGproc}, and spectra were extracted with \texttt{SpectrumExtractor}. To improve SNR, an aperture width of 40 pixels was used for XD mode, and 200 pixels for LR mode. The averaged argon and neon spectra (SNR $>$100 at $\lambda \sim 1~\mu\mathrm{m}$) were combined to form the template for each slit. The XD templates (all orders) are shown in the top panel of \autoref{fig:xd_temp}, and the LR templates in the bottom panel.

During the wavelength calibration step, the pipeline extracts the argon and neon spectra separately, combines them, matches them to the appropriate template, applies the wavelength solution, and saves it to the output file.

\subsection{Flux Calibration}
\label{sec:orgb84d908}
Flux calibration removes wavelength-dependent throughput differences introduced by the telescope and instrument optics, collectively referred to as the \textit{instrument response}. These effects are multiplicative and must be corrected to recover the intrinsic spectrum of the source.

In this version of the pipeline, the TANSPEC instrument response was measured from observations of the A2V-type standard star HD~37725. Spectra of this star were obtained in both LR and XD modes and divided by reference spectra from the CALSPEC catalogue \citep{Bohlin2014TechniquesAR}, produced by the STScI.

\begin{figure*}[htbp]
\centering
\includegraphics[width=.9\linewidth]{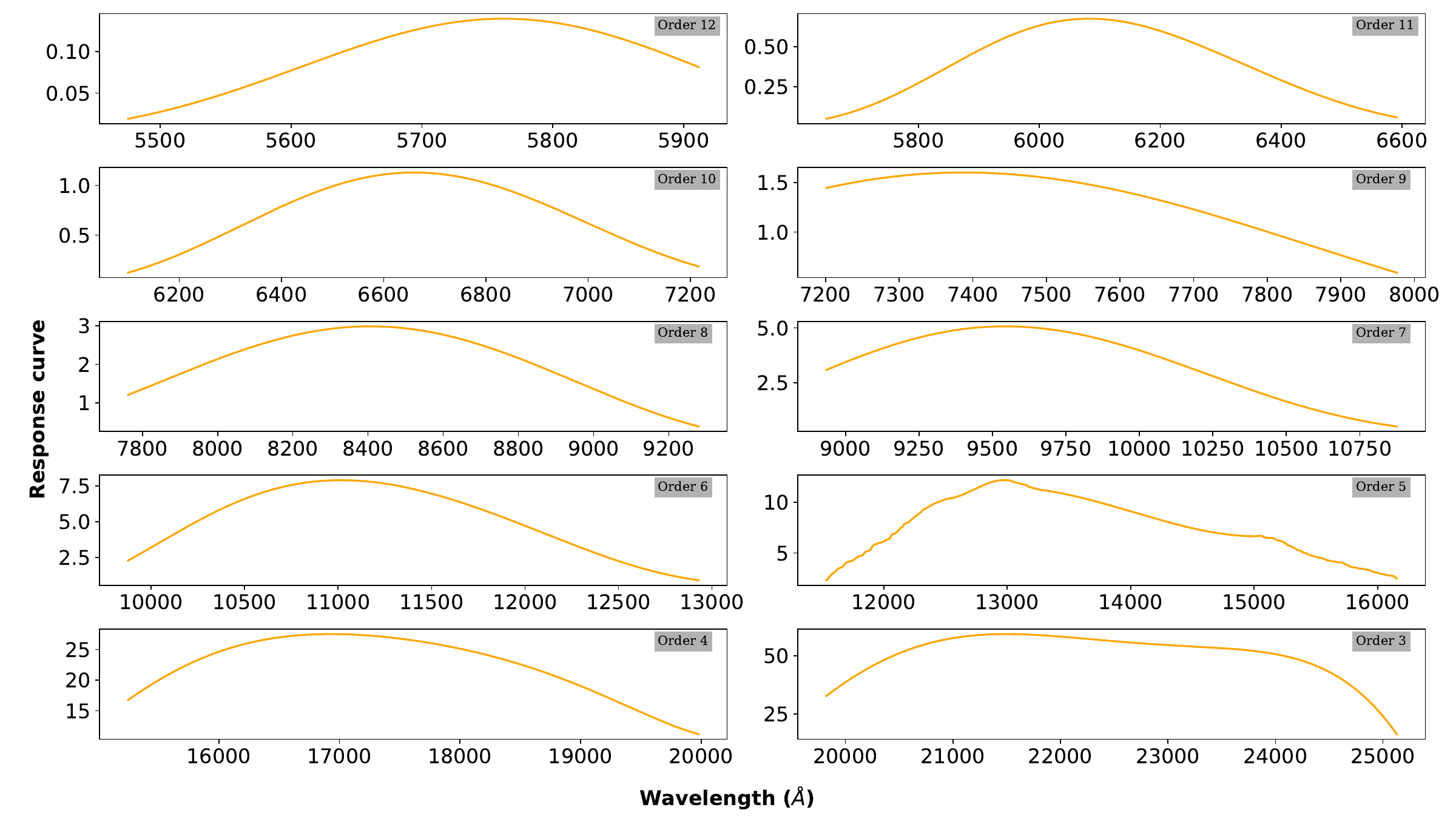}
\includegraphics[width=.9\linewidth]{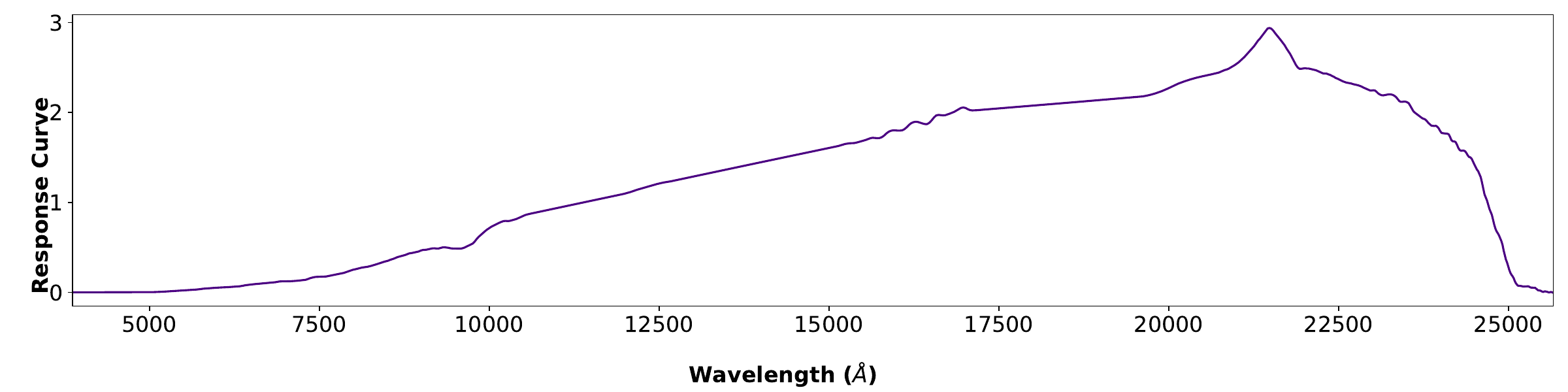}
\caption{\label{fig:xd_resp} Instrument Response curves of for flux calibration. Top panel: XD mode response curves for all spectral orders. Bottom panel: LR mode response curves.}
\end{figure*}

For XD mode, the resulting ratio was smoothed by fitting a polynomial per order. For LR mode, polynomial fitting was unsuitable; instead, the ratio was median-filtered after masking telluric regions to avoid including atmospheric effects in the response. The final response curves for all XD mode orders and LR mode are shown in \autoref{fig:xd_resp}. These response curves are stored with the pipeline, and flux calibration is applied in Task~7 of the reduction process.

\subsection{Variance and S/N}
Along with e\textsuperscript{-}/s slope image, \texttt{HxRGproc} also calculates the variance map while generating the slope image and saves it in \texttt{ext=1}. \texttt{pyTANSPEC} was updated to propagate this variance during spectral extraction, and this is saved along with the extracted spectra. In addition, the pipeline calculates the S/N at the wavelength specified by the user.


\section{Results and Discussion}
\label{sec:org5ae93e5}

We tested the updated package with datasets across multiple nights and observation cycles. As an example, we present the reduced, wavelength- and flux-calibrated spectra of the A2V-type star HD 37725, observed with the slit of width 0.5" and length 60" (S-0.5) in both LR and XD modes. For each frame, slope images were generated using \texttt{HxRGproc}, and the spectra were extracted with \texttt{pyTANSPEC-v1.0}.
For comparison, we also extracted spectra of the same star from QLF. \autoref{fig:hxrg_qlf} compares the two observations. Since \texttt{HxRGproc} applies non-linearity correction, the flux level in the spectra extracted from HxRGproc frames is higher than in those from QLF.

The wavelength- and flux-calibrated spectra are shown in the following figures. \autoref{fig:xd_corr} presents the extracted XD- and LR-mode spectra, along with the CALSPEC standard spectrum of the same source. The observed spectra show good agreement with the standard spectra.

\begin{figure*}[htbp]
\centering
\includegraphics[width=\linewidth]{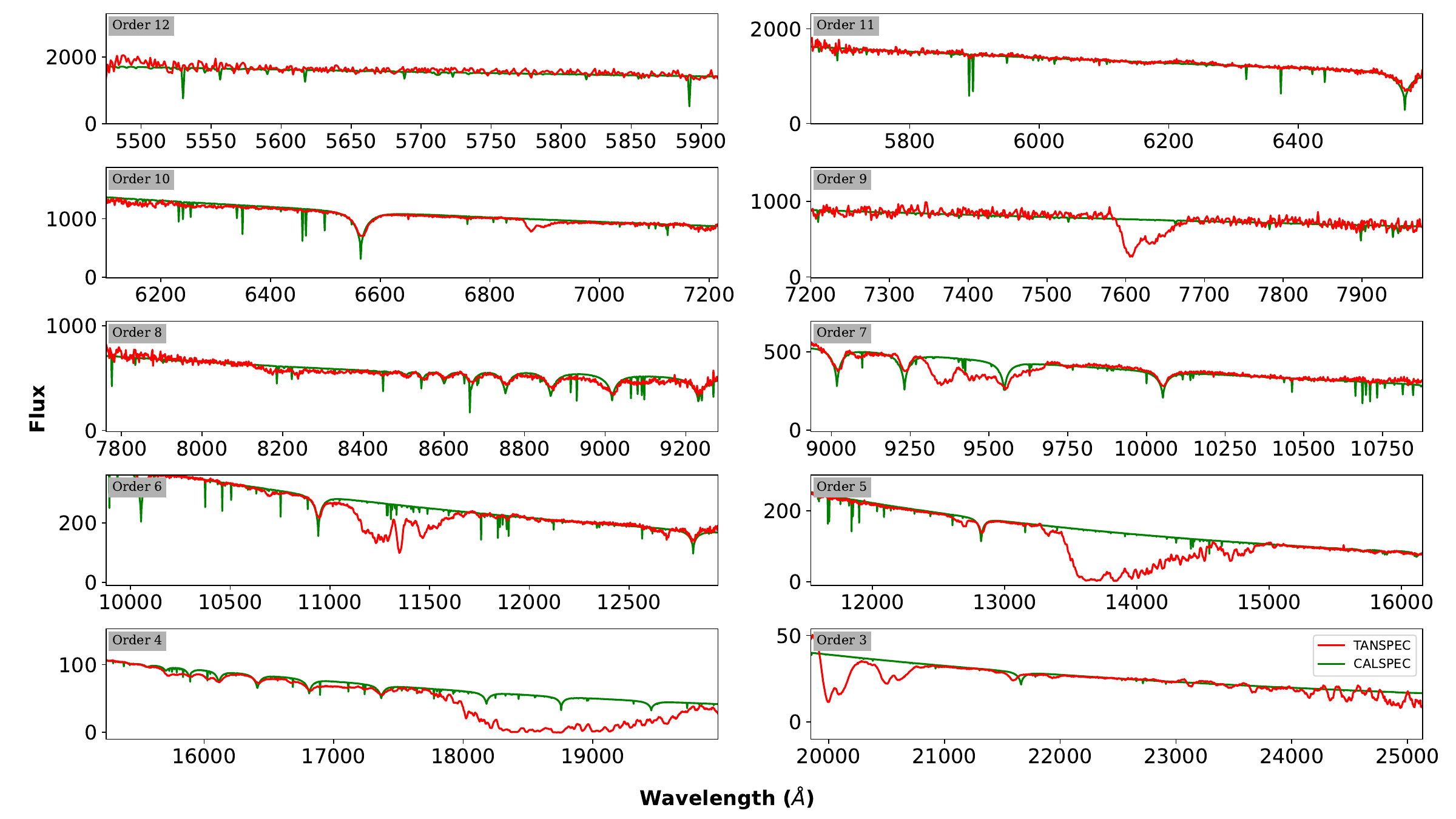}
\includegraphics[width=\linewidth]{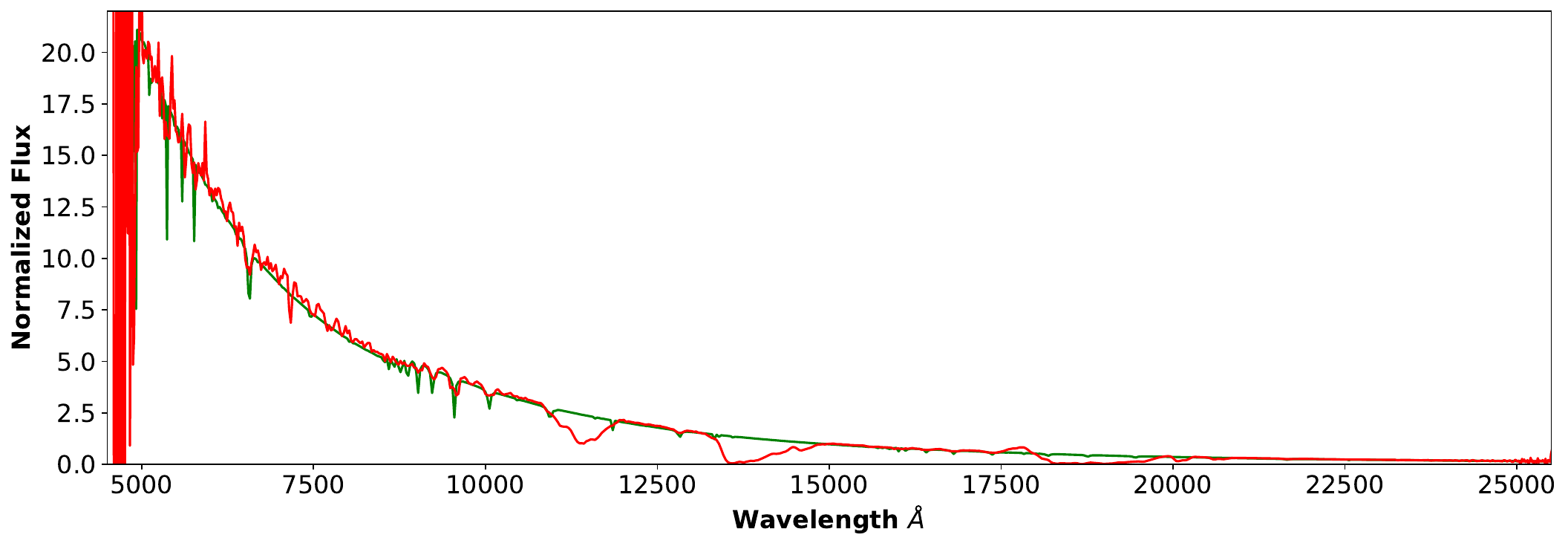}
\caption{\label{fig:xd_corr} Flux-calibrated spectra of HD 37725 from TANSPEC (red) compared with the CALSPEC standard spectrum of the same star (green). The top panel shows the XD-mode spectra, and the bottom panel shows the LR-mode spectra. Both are observed with the slit {S-4.0. Both the spectra follow the spectral energy distribution (SED) of the star}. The data is not telluric corrected.}
\end{figure*}

We tested the pipeline on several GNU/Linux systems, including a Debian 12 machine with a 2.0 GHz Intel Core i3 processor and 12 GB of memory, as well as systems running Ubuntu and Fedora. On all machines, the pipeline required approximately four minutes to complete all tasks (TASK 0 to TASK~7) for the dataset described above. In the new version, wavelength calibration takes less than ten seconds, compared to around one minute in the previous version.

\subsection{Future Improvements}
\label{sec:org503bdd2}

The current version of \texttt{pyTANSPEC} performs only the sum extraction\footnote{The method of spectral extraction that arithmetically adds the counts in the pixels across the aperture at each wavelength without weighting.} of the spectrum. In a future upgrade, we plan to implement optimal extraction\footnote{The method of spectral extraction that combines pixels across the spatial direction using weights based on the spatial profile and noise to maximise the signal-to-noise ratio.}. Currently, the pipeline does not support photometric extraction or telluric correction; we aim to add these capabilities in upcoming versions.

\texttt{pyTANSPEC} is written in a modular framework and can be adapted to extract spectra from any ground-based NIR instrument by providing appropriate aperture trace files, wavelength calibration templates, and flux calibration response functions. For example, uTIRSPEC, mounted on the 2 m Himalayan Chandra Telescope \citep{ninan14:_tirsp}, was recently upgraded to TIRSPEC by replacing its HAWAII-1 PACE array with an H1RG detector (Reji et al., in preparation). We plan to apply \texttt{pyTANSPEC} to uTIRSPEC in the near future.

\section{Summary}
\label{sec:org8c90224}

This work presents the upgraded versions of the \texttt{HxRGproc} and \texttt{pyTANSPEC} packages, which can be used to clean and extract spectra obtained with both resolution arms of TANSPEC. \texttt{HxRGproc} is a Python package designed to generate slope images from NDR frames captured with the H2RG detector of TANSPEC. The package can reduce cosmic rays, correct bias, and correct non-linearity. An improvement of $\sim 13\%$ is observed in the count level after the non-linearity correction.

\texttt{pyTANSPEC-v1.0} is an upgraded version of \texttt{pyTANSPEC-v0.0.1}, originally developed to extract XD spectra from slits S-0.5 and S-1.0. The new version supports the extraction, wavelength calibration, and flux calibration of spectra in both LR and XD modes for all slits. The extraction procedure is identical for LR and XD modes. For wavelength calibration, the updated pipeline matches observed lamp spectra with pre-saved template spectra for each slit. This approach is both faster, more robust and more accurate than the previous method of re-identifying spectral lines. Also, the cycle-to-cycle shift in the pixel position does not affect the wavelength calibration in the new method. On a typical user computer, the pipeline requires approximately five minutes to produce wavelength- and flux-calibrated spectra for LR mode and all orders of XD mode.

.

\section*{Acknowledgements}
VR, JPN and DKO acknowledge the support of the Department of Atomic Energy, Government of India, under Project Identification No. RTI 4002. We thank the members of the IR astronomy group at TIFR for their support. The authors are grateful to the technicians and staff of the 3.6 m DOT, Devasthal, and ARIES, Nainital, for their assistance during the observations. VR thanks Priyankush Ghosh, Priyanka Bhagel, Goldy Ahuja, Shridharan Balakrishnan, Tarak Chand, and others for testing the pipeline before its public release. VR and JPN are grateful to Asish Devaraj and Geetha Rangwal for setting up \texttt{HxRGproc} on the TANSPEC server. This project is supported in part by a generous donation (from the Murty Trust) aimed at enabling advances in astrophysics through the use of machine learning. Murty Trust, an initiative of the Murty Foundation, is a not-for-profit organisation dedicated to preserving and celebrating culture, science, and knowledge systems born out of India. Mrs Sudha Murty and Mr Rohan Murty head the Murty Trust.

\bibliography{Bibliography}
\bibliographystyle{aa}
\end{document}